\documentclass{article}
\usepackage[utf8]{inputenc}
\usepackage{graphicx}
\usepackage{amsmath}
\usepackage{amssymb}
\usepackage{xcolor}
\usepackage{url}
\usepackage[margin=1.0in]{geometry}
\usepackage{algorithm}
\usepackage[noend]{algpseudocode}
\usepackage{comment}
\usepackage{placeins}
\usepackage{authblk}
\usepackage{afterpage}
\usepackage{amsthm}
\usepackage{soul}
\usepackage{siunitx}
\usepackage{tabularx}
\usepackage{subfigure}
\usepackage{afterpage}
\usepackage{rotating}  
\usepackage{fancyhdr}
\usepackage{tabularx}
\usepackage{pgfplots}
\usepackage{caption}
\usepackage{babel,blindtext}
\usepackage{graphicx}

\usepackage{multirow}

\usepackage{float}
\newfloat{algorithm}{t}{lop}

\title{Multi-task graph neural networks for simultaneous prediction of global and atomic properties in ferromagnetic systems}
\author{Massimiliano Lupo Pasini$^{1*}$, Pei Zhang$^1$, Samuel Temple Reeve$^1$, Jong Youl Choi$^2$}
\date{}

\begin{document}

\maketitle

\noindent {$^1$ Oak Ridge National Laboratory, Computational Sciences and Engineering Division, Oak Ridge, TN 37831, USA}\\
{$^2$ Oak Ridge National Laboratory, Computer Science and Mathematics Division, Oak Ridge, TN, 37831, USA}\\
{$^*$ \texttt{lupopasinim@ornl.gov}}

\begin{abstract}
    We introduce a multi-tasking graph convolutional neural network, HydraGNN, to simultaneously predict \textit{both global and atomic} physical properties and demonstrate with ferromagnetic materials. We train HydraGNN on an open-source ab initio density functional theory (DFT) dataset for iron-platinum (FePt) with a fixed body centered tetragonal (BCT) lattice structure and fixed volume to simultaneously predict the mixing  enthalpy (a  global  feature  of  the system), the atomic charge transfer, and the atomic magnetic moment across configurations that span the entire compositional range.  
    By taking advantage of underlying physical correlations between material properties, multi-task learning (MTL) with HydraGNN provides effective training even with modest amounts of data. Moreover, this is achieved with just one architecture instead of three, as required by single-task learning (STL). 
    The first convolutional layers of the HydraGNN architecture are shared by all learning tasks and extract features common to all material properties. The following layers discriminate the features of the different properties, the results of which are fed to the separate heads of the final layer to produce predictions. 
    Numerical results show that HydraGNN effectively captures the relation between the configurational entropy and the material properties
    over the entire compositional range. Overall, the accuracy of simultaneous MTL predictions is comparable to the accuracy of the STL predictions.
    In addition, the computational cost of training HydraGNN for MTL is much lower than the original DFT calculations and also lower than training separate STL models for each property. 
\end{abstract}

{\footnotesize \noindent This manuscript has been authored in part by UT-Battelle, LLC, under contract DE-AC05-00OR22725 with the US Department of Energy (DOE). The US government retains and the publisher, by accepting the article for publication, acknowledges that the US government retains a nonexclusive, paid-up, irrevocable, worldwide license to publish or reproduce the published form of this manuscript, or allow others to do so, for US government purposes. DOE will provide public access to these results of federally sponsored research in accordance with the DOE Public Access Plan (\url{http://energy.gov/downloads/doe-public-access-plan}).}

%

\section{Introduction}
Material discovery and design of new materials relies heavily on predicting material properties directly from their atomic structure. 
There are many physics-based computational approaches to model and predict the behavior of materials at the atomic scale from first principles, such as density functional theory (DFT) \cite{Hoenberg, Kohn}, quantum Monte Carlo (QMC) \cite{qmc, Hammond} and \textit{ab initio} molecular dynamics (MD) \cite{car-parrinello, marx}.
While these methods have been instrumental in predictive materials science, 
they are extremely computationally expensive. 
The advent of data-driven modeling techniques has provided new methodologies to produce inexpensive and accurate predictions of material properties which helps enable rapid screening of large material search spaces to select potential material candidates with desirable properties \cite{gaultois, lu, gomez, xue}. 
Among all data driven models, deep learning (DL) models have the highest potential to accurately represent complex relations between input features and target quantities, but require large volumes of data to attain high accuracy. Data collected in material science is most often small in volume due to expensive experiments and time-consuming numerical simulations, which challenges, but does not preclude, the use of DL models in the field \cite{choudhary2021}. Recent efforts have proposed transfer learning) \cite{hutchinson} and injecting the model with pre-existing physical knowledge \cite{RAISSI2019686, purja, karniadakis} to overcome this data
constraint. 

Multi-task learning (MTL) consists of using one DL model to perform several training tasks at the same time \cite{caruana}. All training tasks mutually influence each other, acting as inductive biases and thus improving each other's predictive performance. When multiple target quantities are correlated with each other, MTL can be used to identify and learn features that are common to all the quantities of interest and transfer knowledge through these common features from one quantity to another. This not only helps to counteract the challenge of data scarcity, but also significantly reduces the computational effort with respect to single-task learning (STL) because only one MTL model is used to predict all properties simultaneously, rather than several distinct STL neural networks. 
MTL is a specific type of physics informed DL when applied to material science data, because it leverages correlations between multiple material properties dictated by the physics \cite{collobert, ramsundar, pasini}.  

Graph convolutional neural network (GCNN) models are a type of transfer learning (TL) that aims at extracting information from local interactions between nodes of graph, and transfers the learnt interactions from one local neighborhood to another, to alleviate the computational burden of DL training. Currently, GCNN models are extensively used in material science to predict material properties from atomic information by directly mapping the atomic structure input to graphs, with atoms as graph nodes and chemical bonds as edges \cite{cgcnn, megnet}. Bond angles \cite{alignn2021} and crystallographic information \cite{icgcnn2020} have also been directly included in GCNN models to improve predictive accuracy. GCNNs not only reduce the cumbersome and expensive data pre-processing, but can also naturally transfer the learning across lattices of different structures and sizes.

Recently, MTL has been combined with GCNN models to strengthen the TL property along with the ability to inject physics knowledge \cite{kim1} in the DL model. In particular, the MT-CGCNN model \cite{mtcgcnn} has been trained on DFT-calculated ordered compounds to simultaneously predict multiple total material properties including mixing enthalpy, Fermi energy, and band gap. Separately, GCNN models have been used to predict per-atom quantities, as with GNNFF which directly predicts atomic forces for molecular dynamics~\cite{gnnff2020}. However, to the best of our knowledge, no existing work in the literature has used multi-task GCNNs to simultaneously predict both global and atomic material properties. In addition, existing approaches that combine MTL with GCNN focus on one specific graph convolutional layer, without allowing the user to flexibly switch among different aggregation policies to customize the convolutional kernel to the nature of the data.

\begin{figure}[htb!]
\centering
\includegraphics[width=15cm]{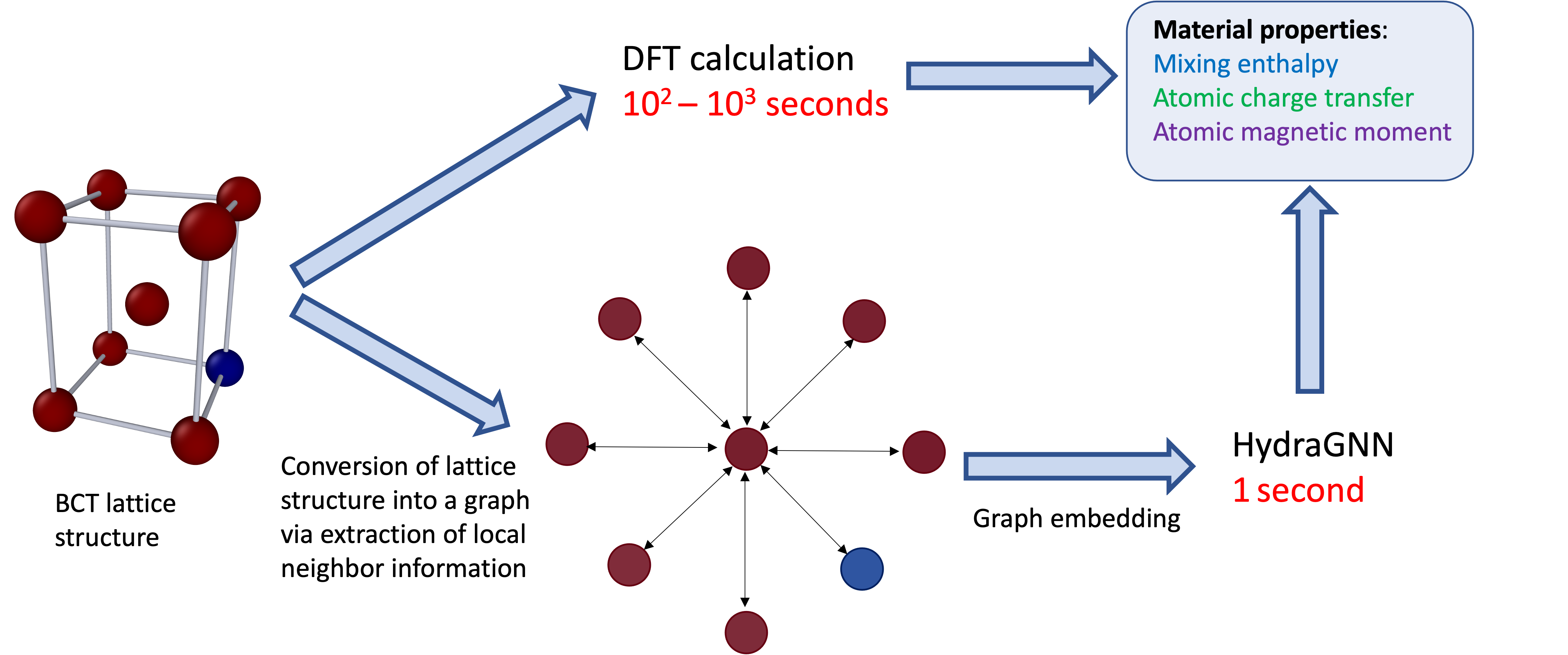}
\caption{ Computational workflow that compares the standard procedure to predict material properties with DFT calculations and the DL-driven methodology that uses the lattice structure as input for HydraGNN to estimate the material properties. Once HydraGNN is trained, it is much faster than DFT.}
\label{fig:HydraGNN-High-level-overview}
\end{figure}

We present a novel multi-tasking GCNN, HydraGNN, to simultaneously predict multiple physical properties and demonstrate its predictive ability for binary solid solution magnetic alloys. We consider mixing enthalpy, atomic charge transfer, and atomic magnetic moment as target material properties. 
Once the HydraGNN model is trained, it can simultaneously produce accurate predictions of multiple material properties substantially faster than DFT calculations. As shown in Figure~\ref{fig:HydraGNN-High-level-overview}, HydraGNN inputs each atomic structure, converts it into a graph, and predicts the same properties as DFT with a GCNN model.
We train HydraGNN on open-source DFT data for iron-platinum (FePt) \cite{FePt} with a fixed body centered tetragonal (BCT) structure and fixed volume, generated with the LSMS-3 code \cite{lsms-code}. 
Numerical results show that HydraGNN learns the dependency of the three material properties with respect to the configurational entropy over the entire compositional range, and the accuracy of the predictions produced with HydraGNN is comparable to the DFT calculations used to train the model (for the system studied).

\section{HydraGNN architecture}
\begin{figure}[htb!]
\centering
\includegraphics[width=15cm]{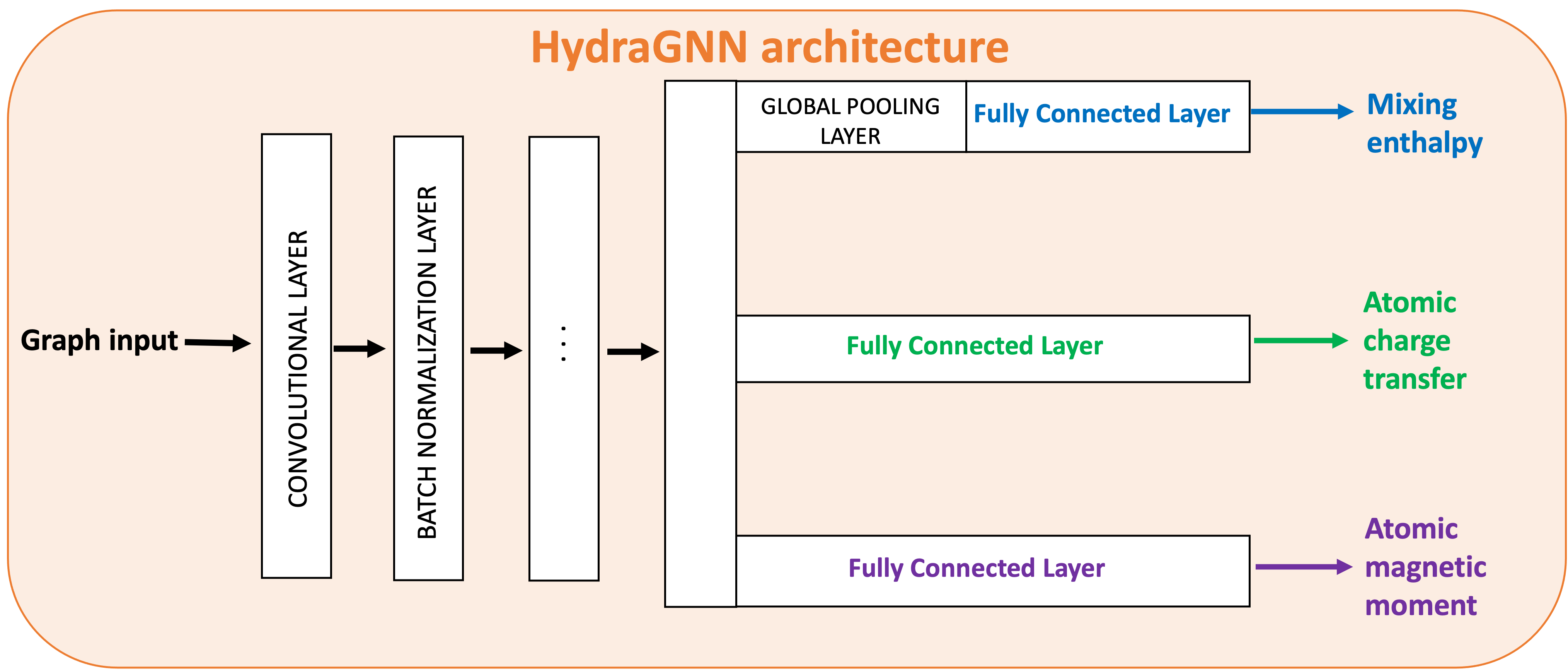}\\
\caption{HydraGNN architecture when used as a surrogate model for DFT calculations of mixing enthalpy, atomic charge transfer, and atomic magnetic moment.}
\label{fig:HydraGNN-architecture}
\end{figure}

HydraGNN directly inputs atomic structure and converts it into a graph, where atoms are interpreted as nodes and interatomic bonds are interpreted as edges, and outputs total (graph-level) and atomic (node-level) physical properties. 
The architecture of HydraGNN is characterized by two sets of layers: the first set of layers learn features that are common to all the material properties and the last set of layers are separated into multiple heads to learn features that are specific to each material property, shown schematically in Figure~\ref{fig:HydraGNN-architecture}. 
The shared graph convolutional layers are used to extract common relevant features from pairwise neighbor interactions and, through multiple layers, also represent many-body interactions. The following separate layers are fully connected and learn mappings between extracted features and the physical properties of interest.

We have implemented HydraGNN using \texttt{Pytorch} \cite{pytorch2019, pytorch} as both a robust NN library, as well as a performance portability layer for running on multiple hardware architectures. This enables HydraGNN to run on CPUs and GPUs, from laptops to supercomputers, including ORNL's Summit. The \texttt{Pytorch Geometric} \cite{fey_2019, torch_geometric} library built on \texttt{Pytorch} is particularly important for our work and enables many GCNN models to be used interchangeably. HydraGNN is openly available on GitHub \cite{hydragnn}.

\subsection{Graph convolutional layers}
{
A graph $G$ is usually represented in mathematical terms as 
\begin{equation}
G = (V, \mathcal{E})
\end{equation}
where $V$ represents the set of nodes and $\mathcal{E}$\ represents the set of edges between these nodes \cite{ja_bondy_usr_murty_graphs_nodate}. An edge $(u,v) \in \mathcal{E}$ connects nodes $u$ and $v$, where ${u,v}\in V$, $\ \mathcal{E} \in V \times V$. 
The topology of a graph can be described through the \textit{adjacency matrix}, $A$, an $N\times N$ square matrix where $N$ is the number of nodes in the graph, whose entries are associated with edges of the graph according to the following rule: 
\begin{equation}
\begin{cases}
A[u,v] = 1 \quad \text{iff} \quad (u,v)\in \mathcal{E} \\
A[u,v] = 0 \quad \text{otherwise}.
\end{cases}
\end{equation}
The degree of a node $u \in V$ is defined as: 
\begin{equation}
d_u = \sum_{v\in V} A[u,v]
\end{equation}
and represents the number of edges connected to a node.}
Every node $u$ is represented by a $a$-dimensional feature vector $\bold x \in \mathbb{R}^a$ containing the embedded nodal properties and also a label vector $\bold y \in \mathbb{R}^b$ in tasks related to node-level predictions. 
In order to take advantage of the topology of the graph, many DL models include both the number of neighbors per node, as well as the length of each edge between nodes.
 
GCNNs embed the interactions between nodes without increasing the size of the input by representing the local interaction zone as a hyperparameter that cuts-off the interaction of a node with all the other nodes outside a prescribed local neighborhood. This is identical to the approximation made by many atomic simulation methods, including the LSMS-3 code used to generate the DFT training data, which ignore interactions outside a given cutoff range. 
GCNNs \cite{GNNpaper, GCNNpaper} are DL models based on a message-passing framework, a procedure that combines the knowledge from neighboring nodes, which in our applications maps directly to the interactions of an atom with its neighbors.

The typical GCNN architecture is characterized by three different types of hidden layers: graph convolutional layers, graph pooling layers, and fully connected layers. 
The convolutional layers represent the central part of the architecture and their functionality is to transfer feature information between adjacent nodes (in this case atoms) iteratively.
In the $k$th convolutional layer ($k=0,1,\ldots, K$), message passing is performed in sequential with the following operations:
\begin{enumerate}
  \item Aggregate information from neighbors: the node $u$ collects the hidden embedded features of its neighbors $N(u)$ as well as the information on the edges (if available) via an aggregation function:
\begin{equation}
\bold {h}_{N(u)}^{k+1}=\mathrm{AGGREGATE}\left(\bold {m}_v^{k}, \forall v\in N(u)\right),
\end{equation}
where $\bold {m}_v^k=\mathrm{MESSAGE}(\bold {h}_v^k,\bold {h}_{e_{uv}}^k)$ is a message obtained from neighboring node $v$ and the edge $e_{uv}$ that connects them.
The vector $\bold {h}_v^k$ ($\bold {h}_v^k\in \mathbb{R}^{p_k}$) is the embedded hidden feature vector of node $v$ in the $k$th convolutional layer. When $k=0$, the hidden feature vector is the input feature vector, $\bold{h}_v^0=\bold x$. 
  \item Update hidden state information: with $\bold {h}_{N(u)}^{k+1}$ collected, the nodal feature of node $u$ is updated as in:
\begin{equation}
\bold {h}_u^{k+1} = \mathrm{UPDATE}\left( \bold {h}_u^k, \bold {h}_{N(u)}^{k+1}\right)
\end{equation}
where $\mathrm{UPDATE}$ is a differentiable function which combines aggregated messages $\bold {h}_{N(u)}^{k+1}$ from neighbors of node $u$ with its nodal features $\bold {h}_u^k$ from the previous layer $k$.
\end{enumerate}

Through consecutive steps of message passing, the graph nodes gather information from nodes that are further and further away. 
The type of information passed through a graph structure can be either related to the topology of the graph or features assigned to the nodes. 
An example of a topological information is the node degree, whereas an example of  nodal feature in the context of this work is the proton number of the atom.
A variety of GCNNs, e.g., principal neighborhood aggregation (PNA) \cite{corso_principal_2020}, crystal GCNN (CGCNN) \cite{cgcnn} and GraphSAGE \cite{hamilton2017inductive}, have been developed, differing in the definitions of functions $\mathrm{AGGREGATE, MESSAGE, UPDATE}$ for message passing.
One simple example of the function combination is:
\begin{equation}
\begin{aligned}
\bold {h}_{N(u)}^{k+1}= \bold{W}_{neighborhood}^{(k+1)}\sum_{v\in N(u)}\bold {h}_v^k+\bold {b}^k, \\
\bold {h}_u^{k+1} = \sigma\left(\bold{W}_{self}^{(k+1)}\bold {h}_u^k + \bold {h}_{N(u)}^{k+1}\right),
\end{aligned}
\end{equation}\label{eq:conv-example}
where $\bold{W}_{self}^{(k+1)},\bold{W}_{neighborhood}^{(k+1)} \in \mathbb{R}^{p_{k+1}\times p_k}$
are the weights of ($k+1$)th layer of GCNN and $\sigma$ is an activation function (e.g., ReLU) that introduces nonlinearity to the model. 

PNA is used in this work and is one of the convolutional layers available in HydraGNN through Pytorch Geometric; PNA combines multiple aggregating techniques to reduce the risk of classifying two different graphs as identical.
Batch normalizations are performed between consecutive convolutional layers along with a ReLU activation function.
Graph pooling layers are connected to the end of the convolution-batch normalization stack to gather feature information from the entire graph.
Fully connected (FC) layers are positioned at the end of the architecture to take the results of pooling, i.e. extracted features, and provide the output prediction. 

Larger sizes of the local neighborhood lead to a higher computational cost to train the HydraGNN model, as the number of regression coefficients to train at each hidden convolutional layer increases 
proportional to the number of neighbors.
Further details on the behavior of HydraGNN with different sizes of the local neighborhood have been previously reported \cite{lupo_gcnn}. 


\subsection{Multiple heads with fully connected layers for multi-task learning}
MTL 
utilizes a NN to simultaneously predict multiple quantities \cite{caruana} when those predicted quantities are mutually correlated and can act as inductive biases for each other. 
The improvement of an MTL model depends on how strongly the quantities to be predicted are mutually correlated in a particular application. This type of field specific inductive bias has been defined as \textit{knowledge-based}, for which the training of a quantity can benefit from the information contained in the training signal for other quantities. 
Ultimately, MTL allows a direct and automated incorporation of physics knowledge into the model by extracting correlations between multiple quantities, with manual intervention by a domain expert only needed in determining which quantities to use.

In HydraGNN, each predicted quantity is associated with a separate loss function and the global objective function minimized during NN training is a linear combination of these individual loss functions. 
Formally, let $T$ be the total number of physical quantities, or tasks, we want to predict. A single task identified by index $i$ focuses on reconstructing a function $f_i:\mathbb{R}^{a}\rightarrow \mathbb{R}^{b_i}$ defined as
\begin{equation}
\bold{y}_i = f_i(\bold x), \quad i=1,\ldots,T ,
\label{functions}
\end{equation}
where $\bold{x} \in \mathbb{R}^{a}$, $\bold{y}_i \in \mathbb{R}^{b_i}$. 
The MTL makes use of the correlation between the quantities $\bold{y}_i$, where the functions $f_i$ in \eqref{functions} are replaced by a single function $\displaystyle \hat{f}:\mathbb{R}^a\rightarrow \mathbb{R}^{\sum_i^T b_i}$ that can model all the relations between inputs and outputs as follows: 
\begin{equation}
\begin{bmatrix}
\bold{y}_1 \\ \vdots \\ \bold{y}_T \end{bmatrix} = \hat{f}_{\bold{W}_\mathrm{MTL}}(\bold{x}),
\label{hat_f}
\end{equation}
where $\bold{W}_\mathrm{MTL}$ represents the weights to be learned. 

The global loss function $\ell_\mathrm{MTL}:\mathbb{R}^{N_\mathrm{MTL}}\rightarrow \mathbb{R}^+$ to be minimized in MTL is a linear combination of the loss functions for the single tasks:
\begin{equation}
    \ell_\mathrm{MTL}(\mathbf{W}_\mathrm{MTL}) = \sum_{i=1}^T \alpha_i \lVert \mathbf{y}_{\textrm{predict},i} 
    - \mathbf{y}_i \rVert_2^2,
\label{global_loss}
\end{equation}
where $\mathbf{y}_{\textrm{predict},i}=\hat{f}_{\mathbf{W}_\mathrm{MTL},i}\left(\bold x\right)$ is the vector of predictions for the $i^\textrm{th}$ quantity of interest and $\alpha_i$ (for $i=1,\ldots,T$) are the mixing weights for the loss functions associated with each single quantity. 
The values of the $\alpha_i$'s in Equation \eqref{global_loss} are hyperparameters of the surrogate model and thus can be tuned. In this work we assigned an equal weight to each property being predicted because we are equally interested in all of them
; however, this definition of the loss function enables one to 
modify the values of the $\alpha_i$ 
to purposely favor the training towards one property of interest.

As mentioned above, the multiple quantities in MTL can be interpreted as mutual inductive biases because the error of a single quantity acts as a regularizer with respect to the loss functions of other quantities. For a fair comparison and to determine the benefit of using other tasks as a mutual regularizer, we do not use additional regularizers for the STL training in this work.

Figure~\ref{fig:HydraGNN-architecture} shows the topology of an HydraGNN model for multi-tasking learning to model mixing enthalpy, atomic charge transfer, and atomic magnetic moment. The architecture of HydraGNN for MTL is organized so that the first hidden layers are shared between all tasks, while keeping several task-specific output layers. This approach is known in literature as \textit{hard parameter sharing}.

\section{Solid solution binary alloy dataset}
\label{dataset_section}

\begin{figure}[h]
    \centering
    \includegraphics[width=0.45\textwidth]{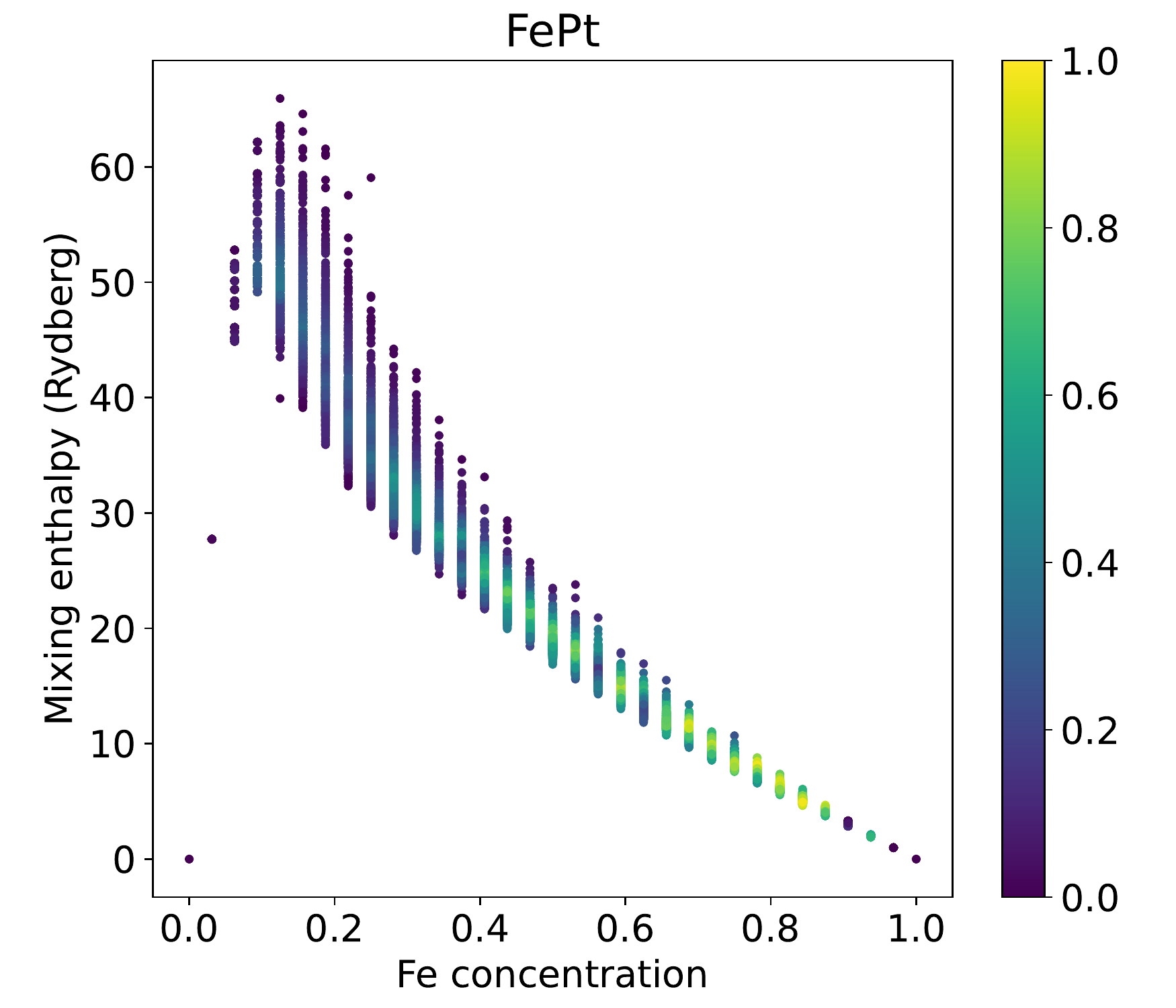}
    \caption{Configurational mixing enthalpy of solid solution binary alloy FePt with BCT structure as a function of Fe concentration. The color map indicates the relative frequency of data.}
    \label{enthalpy}
\end{figure}

In this work we focus on a solid solution binary alloy, where two constituent elements are randomly placed on an underlying crystal lattice. We use a dataset for FePt alloys available through the OLCF Constellation~\cite{FePt} which includes the total enthalpy, atomic charge transfer, and atomic magnetic moment. Each atomic sample has a body centered tetragonal (BCT) structure with a $2 \times 2 \times4$ supercell. The dataset was computed with LSMS-3 \cite{lsms-code}, a locally self-consistent multiple scattering (LSMS) DFT application \cite{eisenbach, lsms}. The dataset was created with fixed volume in order to isolate the effects of graph interactions and graph positions for models such as GCNN. This produces non-equilibrium alloy samples, with non-zero pressure and positive mixing enthalpy, shown as a function of composition in Figure~\ref{enthalpy}.

The input to HydraGNN for each sample includes the three components of the atom position and the proton number. The predicted values include the mixing enthalpy, a single scalar for each sample (graph), as well as the charge transfer and magnitude of the magnetic moment, both scalars per atom (node). Although the magnetic moment is a vector quantity, we treat it as a scalar because all the atomic magnetic moments in the dataset are co-linear 
(all magnetic moments point in the same direction).

The dataset consists of 32,000 configurations out of the $2^{32}$ available, sampled every 3 atomic percent. For this work, if the number of unique configurations for a specific composition is less than 1,000 all those configurations are included in the dataset; for all other compositions, configurations are randomly selected up to 1,000. This results in a final dataset of 28,033 configurations.
In order to ensure each composition is adequately represented in all portions of the dataset, splitting between the training, validation, and test sets is done separately for each composition.

At the ground state, the total enthalpy $H$ of an alloy is
\begin{equation}
    H = \sum_{i=1}^{E} c_i H_i + \Delta H_{\text{mix}},
\end{equation}
where $E$ is the total number of elements in the system, $c_i$ is the molar fraction of each element $i$, $H_i$ is the molar enthalpy of each element $i$, and $\Delta H_{\text{mix}}$ is the mixing enthalpy. 
We predict the mixing enthalpy for each sample by subtracting the internal enthalpy from the DFT computed total enthalpy as a value more relevant to materials science (more directly related to the configuration).
The chemical disorder makes the task of describing the material properties combinatorially complex; this represents the main difference from open source databases that have very broad elemental and structural coverage, but only include ordered compounds \cite{aflow,mp,oqmd}.

The range of values of the mixing enthalpy expressed in Rydberg is $(0.0, 65.92)$, the range of atomic charge transfer in electron charge is $(-5.31, -0.85)$, and the range of atomic magnetic moment in magnetons is $(-0.05, 3.81)$. Since different physical quantities have different units and different orders of magnitude, the inputs and outputs for each quantity are normalized between 0 and 1 across all data.

\section{Numerical results} \label{results}
We present numerical results that predict the mixing enthalpy, atomic charge transfer, and atomic magnetic moment for the binary FePt alloy. Specifically, we compare the predictive performance of multiple separate, single-headed HydraGNN models for STL with multi-headed HydraGNN models for MTL. The output of DFT calculations is considered as the exact reference for the DL model to reconstruct.

\subsection{Training setup}
The architecture of the HydraGNN models has 6 PNA \cite{corso_principal_2020} convolutional layers with 20 neurons per layer. A radius cutoff of 7 angstrom is used to build the local neighborhoods used by the graph convolutional mask. Every learning task is mapped into separate heads where each head is made up of two fully connected layers, with 50 neurons in the first layer and 25 neurons in the second.
The DL models were trained using the Adam method \cite{adam} with a learning rate equal to 0.001, batch sizes of 64, and a maximum number of epochs set to 200. Early stopping is performed to interrupt the training when the validation loss function does not decrease for several consecutive epochs, as this is a symptom that shows further epochs are very unlikely to reduce the value of the loss function.
The training set for each of the NN represents 70\% of the total dataset; the validation and test sets each represent half of the remaining data. As discussed in Section \ref{dataset_section}, compositional stratified splitting was performed to ensure that all the compositions were equally represented across training, validation, and testing datasets.
The training of each DL model was performed on Summit with one model per NVIDIA V100 GPU across two nodes, resulting in ensembles of 12 models per MTL/STL setup discussed in the next section. 

\subsection{Model accuracy and reliability} \label{results}

We compute and analyze not only the accuracy (RMSE) of each model, but also the standard deviation of RMSE for an ensemble of 12 models. This simple metric enables some quantification of the uncertainty associated with each MTL or STL prediction and an understanding of the reliability of the GCNN models.
Note that the RMSE for MTL may be higher compared to STL for two reasons. First, the number of graph convolutional layers and the number of nodes per layer are the same for MTL and STL, but MTL forces these parameters to be shared among the multiple predicted quantities (STL and MTL differ only in the split hidden layers for the heads). Second, MTL introduces an inductive bias in the predictions of a quantity under the influence of other quantities simultaneously predicted. 

\begin{figure}[h]
    \centering
    \includegraphics[width=0.9\textwidth]{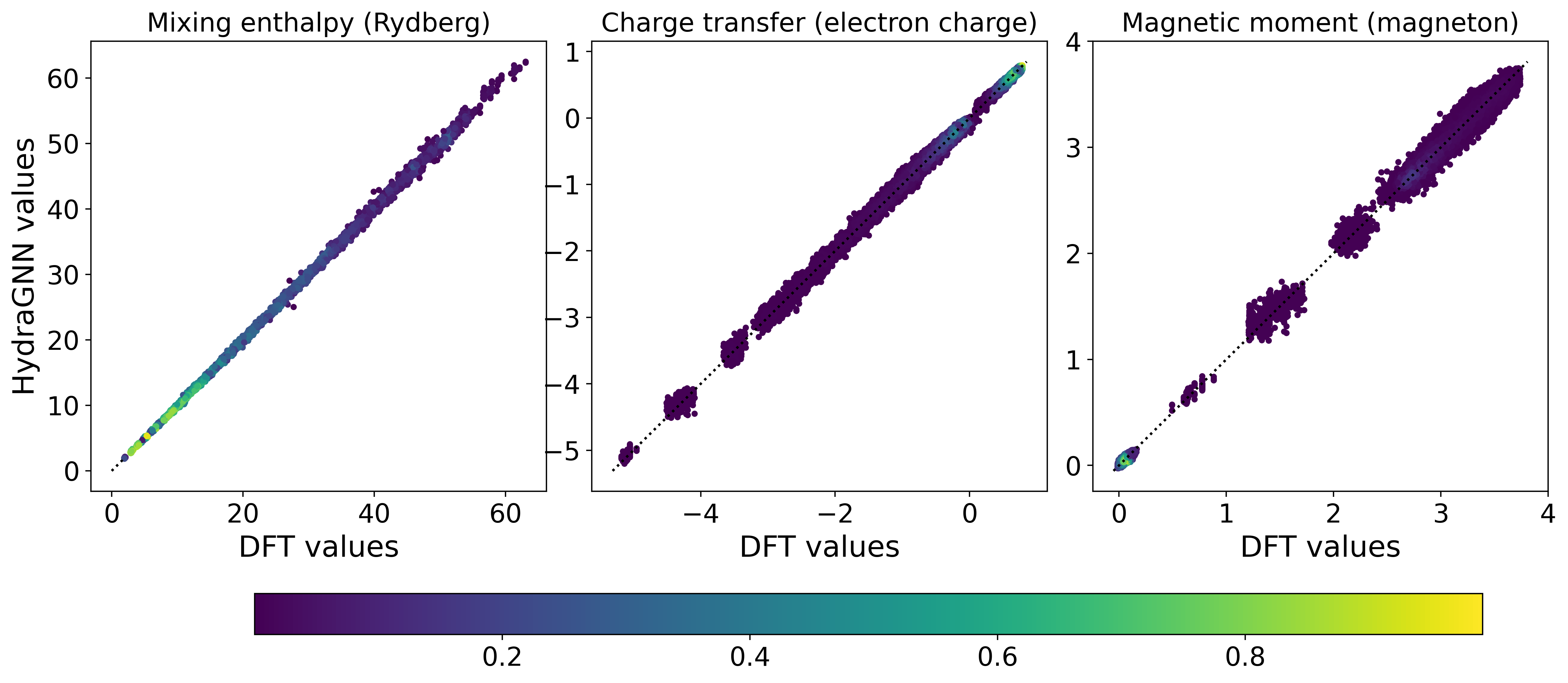}
    \caption{Prediction of FePt test set in MTL, HCM. Scatter plots of mixing enthalpy (left), atomic charge transfer (center), and atomic magnetic moment (right) for multi-task predictions obtained with HydraGNN (y-axis) against DFT calculations (x-axis). The color map indicates the relative frequency of data for each predicted property.}
    \label{multitasking_scatterplot}
\end{figure}

We first show Figure~\ref{multitasking_scatterplot} parity plots for the predictions generated by HydraGNN against the DFT data when MTL is used to simultaneously predict mixing enthalpy, charge transfer, and magnetic moment. HydraGNN with MTL predicts all three properties well for most of the samples over the entire dataset, as shown by the alignment of data near the diagonal. The colormap highlights that more sample points for all three properties are tightly clustered near zero, with larger variations as the property magnitudes increase.
As expected, this non-uniform concentration of values for the target properties across the data affects the predictive performance of the HydraGNN models, with more accurate predictions in regions with higher concentration of data.

\begin{figure}[h]
    \centering
    \includegraphics[width=0.9\textwidth]
    {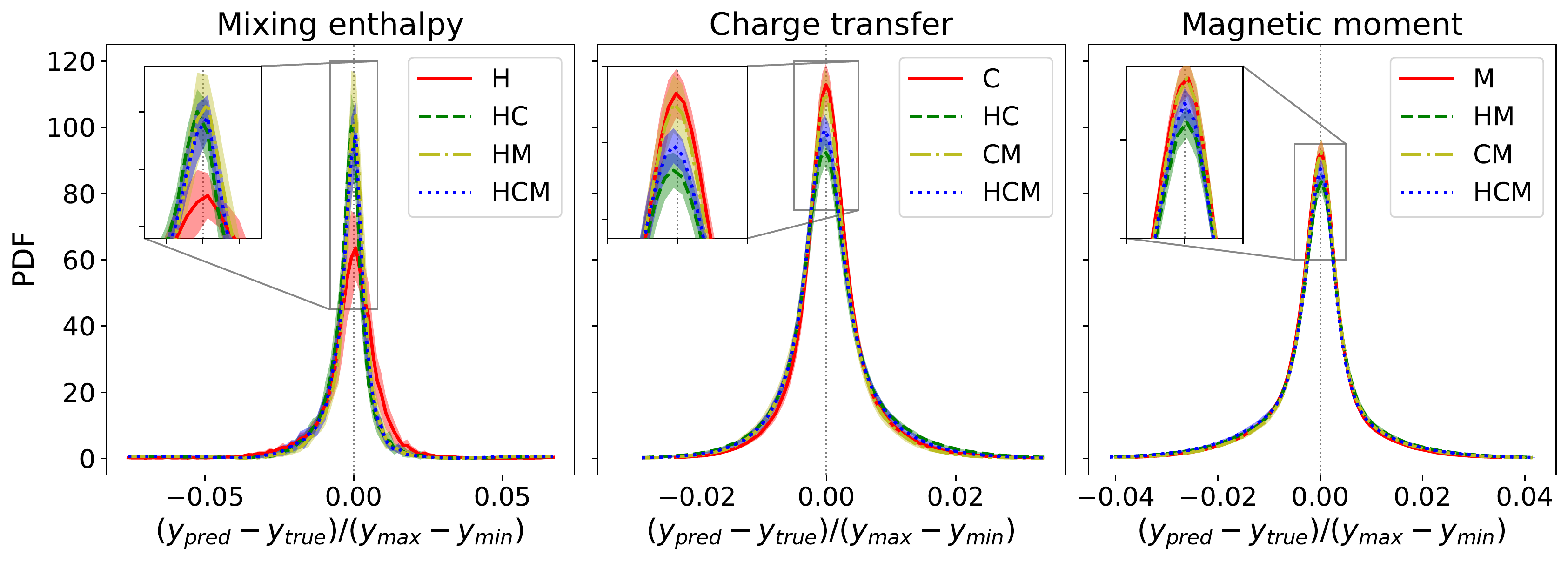}
    \caption{Comparison between prediction error of FePt test set in STL and MTL. PDF profiles of prediction errors for mixing enthalpy (left), atomic charge transfer (center), and atomic magnetic moment (right). The lines are mean values and the filled area indicates the range of standard deviation from 12 independent runs with random initialization. The capital letters are associated with the quantities predicted (H stands for mixing enthalpy, C stands for charge transfer and M stands for magnetic moment).}
    \label{fig:errorpdf}
\end{figure}

In Table~\ref{fept_table}, we compare the predictive accuracy of HydraGNN used to simultaneously predict mixing enthalpy, atomic charge transfer, and atomic magnetic moment with MTL, MTL with each pair of properties, and STL for each individual property. 
The table shows the average root mean-squared error (RMSE) and the standard deviation from 12 independent runs with random initialization on the test set. 
Figure~\ref{fig:errorpdf} shows the probability distribution functions (PDF) of the results from Table~\ref{fept_table} for each combination of properties.
The lines mark the average values and the filled area indicates one standard deviation. Models with a lower average RMSE are interpreted as more accurate and lower standard deviation as more reliable, as the model predictions are similar across different trainings due to more
relevant features extracted that better characterize the dataset.
The results show that adding the magnetic moment as a physical constraint improves the both the predictive accuracy and reliability of MTL models over the STL predicting mixing enthalpy only. Addition or replacement of magnetic momement for charge density somewhat reduces the accuracy or reliability, but much less than using STL. 

These results suggest that the correlation between mixing enthalpy and magnetic moment is stronger than the correlation between mixing enthalpy and charge density. 
This stronger correlation in ferromagnetic materials is supported by DFT calculations and experiments \cite{pinski, hou, marshal}, as well as previous DL studies using multilayer perceptrons for MTL on the same dataset \cite{lupopasini}.

The error distributions for MTL are slightly skewed and not centered at zero because the physical properties operate as mutual inductive biases, but this does not affect the accuracy of MTL. 
While MTL stabilizes the training and prediction of the global graph mixing enthalpy, the same is not true of the atomic-level properties. The STL models for charge density and magnetic moment are slightly more accurate (as discussed at the beginning of the section), but also slightly more reliable for magnetic moment. It is possible that the additional data contained within the per-node properties, more than an order of magnitude above the per-graph property, is sufficient to predict on this dataset with STL alone.

We finally note that while our results for the mixing enthalpy are less accurate than the best reported GCNN results for DFT data ($\sim$0.21 eV/atom in this work versus 0.022 - 0.067 eV/atom \cite{icgcnn2020, alignn2021}). This is partially due to the model used, as PNA does not explicitly include bond angle or crystallographic information, as well as the highly non-equilibrium nature of this dataset (although that the literature results include a much more broad range of chemistries should be taken into account). However, the focus of this work is on the MTL capabilities of HydraGNN and comparisons with STL.


\begin{table}[h]
\centering
  \begin{tabular}{|c|c|c|c|}
    \multicolumn{1}{|c|}{} &
    \multicolumn{3}{c|}{\bf{Test RMSE}} \\
     \multicolumn{1}{|c|}{\bf{Training method}} &
      \multicolumn{1}{c|}{\bf{Mixing enthalpy} } &
      \multicolumn{1}{c|}{\bf{Charge transfer} } &
      \multicolumn{1}{c|}{\bf{Magnetic moment} } \\
MTL, HCM&$\num{7.54e-03}\pm\num{8.70e-04}$ & $\num{6.77e-03}\pm\num{3.59e-04}$ & $\num{1.04e-02}\pm\num{4.94e-04}$\\
MTL, HC&$\num{7.33e-03}\pm\num{4.77e-04}$ & $\num{7.36e-03}\pm\num{3.23e-04}$&-\\
MTL, HM&$\num{6.64e-03}\pm\num{5.08e-04}$&-&$\num{1.02e-02}\pm\num{5.23e-04}$\\
MTL, CM&-&$\num{5.94e-03}\pm\num{3.02e-04}$&$\num{9.30e-03}\pm\num{4.12e-04}$\\
STL, H&$\num{1.02e-02}\pm\num{1.16e-03}$&-&-\\
STL, C&-&$\num{5.94e-03}\pm\num{4.39e-04}$&-\\
STL, M&-&-&$\num{8.77e-03}\pm\num{3.18e-04}$\\
  \end{tabular}
  \caption{Test RMSE of HydraGNN to predict physical properties for an FePt alloy. The training method states whether multi-tasking (MTL) or single-tasking (STL) was used and which quantities were predicted: mixing enthalpy (H), charge transfer (C), and/or magnetic moment (M).}
  \label{fept_table}
\end{table}






\subsection{Model training time}
\begin{figure}[h]
    \centering
    \includegraphics[width=0.5\textwidth]
    {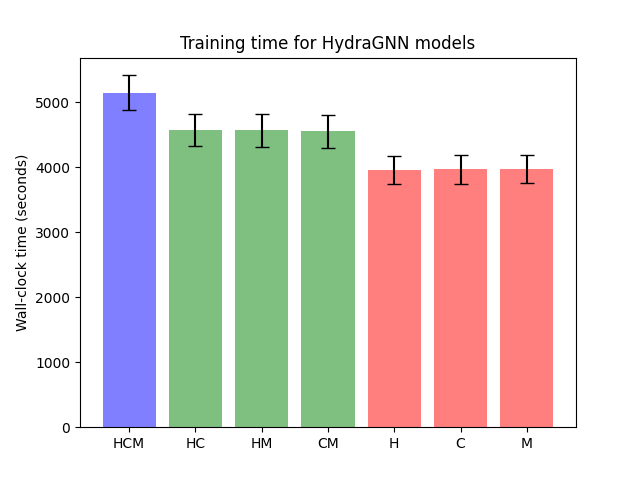}
    \caption{Training time for HydraGNN models in wall-clock seconds to simultaneously predict all three material properties (red), two properties (green) and one property (blue). The bars indicate average training time and the vertical black segments indicate the range of one standard deviation from 12 independent runs with random initialization. Each letter corresponds to one predicted quantity: mixing enthalpy (H), charge transfer (C), and magnetic moment (M).}
    \label{fig:time_bars}
\end{figure}

In Figure~\ref{fig:time_bars}, we report the average wall-clock time and standard deviation to train all the MTL and STL models for the same 12 averaged runs from Section \ref{results} for each combination of properties.
In general, STL takes approximately 10\% less time to train than MTL for two properties, which in turn takes approximately 10\% less than MTL to predict all thee material properties. This is due to the fact that each head introduces additional parameters into the neural network, increasing the total computational cost. However, the training for MTL to simultaneously predict all three material properties is still more than 2.2x faster than to train three separate HydraGNN models using STL.

\section{Conclusion and future work}
We have presented a new DL surrogate model for atomic calculations, HydraGNN, to predict both global and atomic properties, available on GitHub \cite{hydragnn}. The multi-tasking GCNN model was jointly trained on mixing enthalpy, charge transfer, and magnetic moment to take advantage of physical correlations in a highly non-equilibrium DFT dataset. Each predicted quantity acts as a physical constraint on the other quantities, allowing the shared convolutional layers to extract features that describe local interatomic interactions and use them across different material properties.

Although the numerical experiments presented here show that MTL can use strong correlations between physical properties to reduce the uncertainty associated with the DL predictions, imposing too many constraints can be counterproductive. 
Joint training to simultaneously predict too many quantities can lead to a decrease of the predictive power, particularly as the quantities become less correlated. Therefore, the inclusion of different physical properties for use as physical constraints must be determined judiciously. 

The use of HydraGNN reduces the time needed to predict the material properties by a factor of hundreds compared to the original DFT calculations. Moreover, the computational time in training a multi-headed HydraGNN model is comparable to that of a single-headed GCNN model, resulting in further computational savings (near the order of the number of material properties). 

We envisage two primary future applications of HydraGNN. 
First, the surrogate enables extremely fast estimation of atomic properties once the model has been trained, replacing DFT as long as some reduction in accuracy is acceptable. To further improve the usability for this case, we plan to integrate uncertainty quantification methods with HydraGNN to identify situations where the model requires additional training data.
Second, rather than completely replace DFT we intend to integrate physics-based and data-driven models together in scenarios where the trained HydraGNN surrogate model can act as an initial guess for a further refined DFT calculation. In computational workflows such as quantum Monte Carlo, this could substantially reduce the total simulation time without any change in the final accuracy.


\FloatBarrier

\section*{Acknowledgements}
Massimiliano Lupo Pasini thanks Dr. Vladimir Protopopescu and Dr. Markus Eisenbach for their valuable feedback in the preparation of this manuscript.

This work was supported in part by the Office of Science of the Department of Energy and by the Laboratory Directed Research and Development (LDRD) Program of Oak Ridge National Laboratory. 
This research is sponsored by the Artificial Intelligence Initiative as part of the Laboratory Directed Research and Development (LDRD) Program of Oak Ridge National Laboratory, managed by UT-Battelle, LLC, for the US Department of Energy under contract DE-AC05-00OR22725.
This work used resources of the Oak Ridge Leadership Computing Facility and of the Edge Computing program at the Oak Ridge National Laboratory, which is supported by the Office of Science of the U.S. Department of Energy under Contract No. DE-AC05-00OR22725. 

\bibliographystyle{unsrt}
\bibliography{references}

\begin{thebibliography}{10}

\bibitem{Hoenberg}
P.~Hoenber and W.~Kohn.
\newblock Inhomogeneous electron gas.
\newblock {\em Phys. Rev.}, 136:B864–B871, 1964.

\bibitem{Kohn}
W.~Kohn and L.~J. Sham.
\newblock Self-consistent equations including exchange and correlation effects.
\newblock {\em Phys. Rev.}, 140:A1133--A1138, 1965.

\bibitem{qmc}
M.~P. Nightingale and J.~C. Umrigar.
\newblock {\em {Self-Consistent Equations Including Exchange and Correlation
  Effects}}.
\newblock Springer, 1999.

\bibitem{Hammond}
B.~L. Hammond, W.~A. Lester, and P.~J. Reynolds.
\newblock {\em {Monte Carlo Methods in Ab Initio Quantum Chemistry}}.
\newblock Singapore: World Scientific, 1994.

\bibitem{car-parrinello}
R.~Car and M.~Parrinello.
\newblock Unified approach for molecular dynamics and density-functional
  theory.
\newblock {\em Phys. Rev. Lett.}, 55:2471--2474, 1985.

\bibitem{marx}
D.~Marx and J.~Hutter.
\newblock {\em {Ab Initio Molecular Dynamics, Basic Theory and Advanced
  Methods}}.
\newblock Cambridge University Press New York, New York, USA, 2012.

\bibitem{gaultois}
M.~W. Gaultois, A.~O. Oliynyk, A.~Mar, T.~D. Sparks, G.~J. Mulholland, and
  B.~Meredig.
\newblock Perspective: Web-based machine learning models for real-time
  screening of thermoelectric materials properties.
\newblock {\em APL Materials}, 4(053213), May 2016.

\bibitem{lu}
S.~Lu, Q.~Zhou, Y.~Ouyang, Y.~Guo, and J.~Li, Q.;~Wang.
\newblock Accelerated discovery of stable lead- free hybrid organic-inorganic
  perovskites via machine learning.
\newblock {\em Nature Communications}, 9(3405), August 2018.

\bibitem{gomez}
R.~Gómez-Bombarelli.
\newblock Design of efficient molecular organic light-emitting diodes by a
  high-throughput virtual screening and experimental approach.
\newblock {\em Nature Materials}, 15:1120--1127, August 2016.

\bibitem{xue}
D.~Xue, P.~V. Balachandran, J.~Hogden, J.~Theiler, D.~Xue, and T.~Lookman.
\newblock Accelerated search for materials with targeted properties by adaptive
  design.
\newblock {\em Nature Communications}, 7(11241), April 2016.

\bibitem{choudhary2021}
Kamal Choudhary, Brian DeCost, Chi Chen, Anubhav Jain, Francesca Tavazza, Ryan
  Cohn, Cheol WooPark, Alok Choudhary, Ankit Agrawal, Simon J.~L. Billinge,
  Elizabeth Holm, Shyue~Ping Ong, and Chris Wolverton.
\newblock Recent advances and applications of deep learning methods in
  materials science, 2021.

\bibitem{hutchinson}
Maxwell Hutchinson, Erin Antono, Brenna~M. Gibbons, Sean Paradiso, Julia Ling,
  and Bryce Meredig.
\newblock Overcoming data scarcity with transfer learning.
\newblock {\em ArXiv}, abs/1711.05099, 2017.

\bibitem{RAISSI2019686}
M.~Raissi, P.~Perdikaris, and G.E. Karniadakis.
\newblock Physics-informed neural networks: A deep learning framework for
  solving forward and inverse problems involving nonlinear partial differential
  equations.
\newblock {\em Journal of Computational Physics}, 378:686--707, 2019.

\bibitem{purja}
G.~P. Purja, R.~Batra, and Y.~Mishin.
\newblock Physically informed artificial neural networks for atomistic modeling
  of materials.
\newblock {\em Nat. Commun}, 10(2339), 2019.

\bibitem{karniadakis}
G.E. Karniadakis, I.G. Kevrekidis, L.~Lu, and et~al.
\newblock Physics-informed machine learning.
\newblock {\em Nature Reviews Physics}, 3:422--440, May 2021.

\bibitem{caruana}
R.~Caruana.
\newblock Multitask learning: a knowledge-based source of inductive bias.
\newblock {\em Mach. Learn.}, 48:41--48, 1993.

\bibitem{collobert}
R.~Collobert and J.~Weston.
\newblock A unified architecture for natural language processing: Deep neural
  networks with multitask learning.
\newblock {\em Proceedings of the 25th International Conference on Machine
  Learning}, 7:160--167, July 2008.

\bibitem{ramsundar}
B.~Ramsundar, B.~Liu, Z.~Wu, A.~Verras, M.~Tudor, R.~P. Sheridan, and V.~Pande.
\newblock Is multitask deep learning practical for pharma?
\newblock {\em Journal of Chemical Information and Modeling}, (57):2068--2076,
  2017.

\bibitem{pasini}
M.~Lupo~Pasini, Y.~W. Li, J.~Yin, J.~Zhang, K.~Barros, and M.~Eisenbach.
\newblock Fast and stable deep-learning predictions of material properties for
  solid solution alloys.
\newblock {\em J. Phys.: Condens. Matter}, 33(8):084005, December 2020.
\newblock Publisher: IOP Publishing.

\bibitem{cgcnn}
T.~Xie and J.~C. Grossman.
\newblock Crystal graph convolutional neural networks for an accurate and
  interpretable prediction of material properties.
\newblock {\em Phys. Rev. Lett.}, 120(14):145301, April 2018.

\bibitem{megnet}
C.~Chen, W.~Ye, Y.~Zuo, C.~Zheng, and S.~P. Ong.
\newblock Graph networks as a universal machine learning framework for
  molecules and crystals.
\newblock {\em Chem. Mater.}, 31(9):3564--3572, May 2019.

\bibitem{alignn2021}
Kamal Choudhary and Brian DeCost.
\newblock Atomistic line graph neural network for improved materials property
  predictions.
\newblock {\em npj Computational Materials}, 7(1):1--8, 2021.

\bibitem{icgcnn2020}
Cheol~Woo Park and Chris Wolverton.
\newblock Developing an improved crystal graph convolutional neural network
  framework for accelerated materials discovery.
\newblock {\em Phys. Rev. Materials}, 4:063801, Jun 2020.

\bibitem{kim1}
R.~Kim and Y.~Ning.
\newblock Recurrent multi-task graph convolutional networks for covid-19
  knowledge graph link prediction.
\newblock {\em Springer Journal of Communications in Computer and Information
  Science}, September 2021.

\bibitem{mtcgcnn}
S.~Sanyal, J.~Balachandran, N.~Yadati, A.~Kumar, P.~Rajagopalan, S.~Sanyal, and
  P.~P. Talukdar.
\newblock {MT-CGCNN:} integrating crystal graph convolutional neural network
  with multitask learning for material property prediction.
\newblock {\em ArXiv}, abs/1811.05660, 2018.

\bibitem{gnnff2020}
Cheol~Woo Park, Mordechai Kornbluth, Jonathan Vandermause, Chris Wolverton,
  Boris Kozinsky, and Jonathan~P Mailoa.
\newblock Accurate and scalable multi-element graph neural network force field
  and molecular dynamics with direct force architecture.
\newblock {\em arXiv preprint arXiv:2007.14444}, 2020.

\bibitem{FePt}
M.~Lupo~Pasini and M.~Eisenbach.
\newblock {FePt binary alloy with 32 atoms - LSMS-3 data -
  DOI:10.13139/OLCF/1762742}.

\bibitem{lsms-code}
M.~Eisenbach, Y.~W. Li, O.~K. Odbadrakh, Z.~Pei, G.~M. Stocks, and J.~Yin.
\newblock {LSMS}. https://github.com/mstsuite/lsms.

\bibitem{pytorch2019}
A.~Paszke, S.~Gross, F.~Massa, A.~Lerer, J.~Bradbury, G.~Chanan, T.~Killeen,
  Z.~Lin, N.~Gimelshein, L.~Antiga, A.~Desmaison, A.~Kopf, E.~Yang, Z.~DeVito,
  M.~Raison, A.~Tejani, S.~Chilamkurthy, B.~Steiner, L.~Fang, J.~Bai, and
  S.~Chintala.
\newblock Pytorch: An imperative style, high-performance deep learning library.
\newblock In H.~Wallach, H.~Larochelle, A.~Beygelzimer, F.~d\textquotesingle
  Alch\'{e}-Buc, E.~Fox, and R.~Garnett, editors, {\em Advances in Neural
  Information Processing Systems 32}, pages 8024--8035. Curran Associates,
  Inc., 2019.

\bibitem{pytorch}
{PyTorch}.
\newblock \url{https://pytorch.org/docs/stable/index.html}.

\bibitem{fey_2019}
Matthias Fey and Jan~E. Lenssen.
\newblock Fast graph representation learning with {PyTorch Geometric}.
\newblock In {\em ICLR Workshop on Representation Learning on Graphs and
  Manifolds}, 2019.

\bibitem{torch_geometric}
{PyTorch Geometric}.
\newblock \url{https://pytorch-geometric.readthedocs.io/en/latest/}.

\bibitem{hydragnn}
Massimiliano Lupo~Pasini, Samuel~T. Reeve, Pei Zhang, and Jong~Youl Choi.
\newblock {HydraGNN}.
\newblock [Computer Software] \url{https://doi.org/10.11578/dc.20211019.2}, oct
  2021.

\bibitem{ja_bondy_usr_murty_graphs_nodate}
U.S.R.~Murty J.A.~Bondy.
\newblock Graphs and subgraphs.
\newblock In {\em Graph theory with applications}. North-Holland.

\bibitem{GNNpaper}
F.~Scarselli, M.~Gori, A.~C. Tsoi, M.~Hagenbuchner, and G.~Monfardini.
\newblock The graph neural network model.
\newblock {\em IEEE Transactions on Neural Networks}, 20(1):61--80, 2009.

\bibitem{GCNNpaper}
M.~Defferrard, X.~Bresson, and P.~Vandergheynst.
\newblock Convolutional neural networks on graphs with fast localized spectral
  filtering.
\newblock In D.~Lee, M.~Sugiyama, U.~Luxburg, I.~Guyon, and R.~Garnett,
  editors, {\em Advances in Neural Information Processing Systems}, volume~29.
  Curran Associates, Inc., 2016.

\bibitem{corso_principal_2020}
Gabriele Corso, Luca Cavalleri, Dominique Beaini, Pietro Liò, and Petar
  Veličković.
\newblock Principal neighbourhood aggregation for graph nets.
\newblock {\em arXiv:2004.05718 [cs, stat]}, December 2020.
\newblock arXiv: 2004.05718.

\bibitem{hamilton2017inductive}
William~L Hamilton, Rex Ying, and Jure Leskovec.
\newblock Inductive representation learning on large graphs.
\newblock In {\em Proceedings of the 31st International Conference on Neural
  Information Processing Systems}, pages 1025--1035, 2017.

\bibitem{lupo_gcnn}
M.~Lupo~Pasini, M.~Bur\^cul, S.~T. Reeve, M.~Eisenbach, and S.~Perotto.
\newblock Fast and accurate predictions of total energy for solid solution
  alloys with graph convolutional neural networks.
\newblock {\em Springer Journal of Communications in Computer and Information
  Science}, September 2021.

\bibitem{eisenbach}
M.~Eisenbach, J.~Larkin, J.~Lutjens, S.~Rennich, and J.~H. Rogers.
\newblock {GPU} acceleration of the locally self-consistent multiple scattering
  code for first principles calculation of the ground state and statistical
  physics of materials.
\newblock {\em Comput. Phys. Commun.}, 211:2--7, 2017.

\bibitem{lsms}
Y.~Wang, G.~M. Stocks, W.~A. Shelton, D.~M.~C. Nicholson, Z.~Szotek, and W.~M.
  Temmerman.
\newblock Order-{N} multiple scattering approach to electronic structure
  calculations.
\newblock {\em Phys. Rev. Lett.}, 75:2867--2870, 1995.

\bibitem{aflow}
S.~Curtarolo, W.~Setyawan, G.~L.~W. Hart, M.~Jahnatek, R.~V. Chepulskii, R.~H.
  Taylor, S.~Wang, J.~Xue, K.~Yang, O.~Levy, M.~J. Mehl, H.~T. Stokes, D.~O.
  Demchenko, and D.~Morgan.
\newblock {AFLOW}: {An} automatic framework for high-throughput materials
  discovery.
\newblock {\em Comput. Mater. Sci.}, 58:218--226, June 2012.

\bibitem{mp}
A.~Jain, S.~Ping Ong, G.~Hautier, W.~Chen, W.~D. Richards, S.~Dacek, S.~Cholia,
  D.~Gunter, D.~Skinner, G.~Ceder, and K.~A. Persson.
\newblock Commentary: {The} materials project: {A} materials genome approach to
  accelerating materials innovation.
\newblock {\em APL Mater.}, 1(1):0--11, 2013.

\bibitem{oqmd}
J.~E Saal, S.~Kirklin, M.~Aykol, B.~Meredig, and C.~Wolverton.
\newblock Materials design and discovery with high-throughput density
  functional theory: the {Open} {Quantum} {Materials} {Database} ({OQMD}).
\newblock {\em JOM}, 65(11):1501--1509, November 2013.

\bibitem{adam}
Diederik~P. Kingma and Jimmy Ba.
\newblock Adam: a method for stochastic optimization.
\newblock {\em arXiv:1412.6980 [cs]}, January 2017.
\newblock arXiv: 1412.6980.

\bibitem{pinski}
F.~J. Pinksi, J.~B. Staunton, S.~S.~A. Razee, and D.~D. Johnson.
\newblock Magnetism in alloys.
\newblock {\em Mater. Charact.}, pages 1--28, 2002.

\bibitem{hou}
T.~P. Hou, K.~M. Liu, M.~J. Peet, C.~N. Hulme-Smith, L.~Guo, and L.~Zhang.
\newblock Magnetism and high magnetic-field-induced stability of alloy carbides
  in fe-based materials.
\newblock {\em Sci. Rep.}, 8(3049), 2018.

\bibitem{marshal}
A.~Marshal, K.~G. Pradeep, D.~Music, L.~Wang, O.~Petracic, and J.~M. Schneider.
\newblock Combinatorial evaluation of phase formation and magnetic properties
  of femncocral high entropy alloy thin film library.
\newblock {\em Sci. Rep.}, 9(7864), 2019.

\bibitem{lupopasini}
M.~Lupo Pasini, J.~Yin, Y.~W. Li, and M.~Eisenbach.
\newblock A scalable constructive algorithm for the optimization of neural
  network architectures.
\newblock {\em Parallel Comput.}, 104--105:102788, 2021.

\end{thebibliography}

\end{document}